\documentclass[aps,prl,showpacs,twocolumn,groupedaddress,amssymb]{revtex4}

\usepackage{graphicx}
\usepackage{dcolumn}

\begin{document}

\title{Soft X-ray Amplification by an intense laser}

\author{S. Son}
\affiliation{169 Snowden Lane, Princeton, NJ, 08540}

\begin{abstract}

A new  x-ray amplification mechanism is considered in an interaction between a x-ray  and an  intense visible-light laser in a plasma.   
In normal circumstances,  the x-ray amplification from this type of the physical processes is  implausible because  
its  phase,   which is the \textbf{non-resonant} perturbation,
is not suitable for transferring the energy from the visible light laser to the x-ray. 
As proposed in this paper, this undesirable  phase can be manipulated  into the desirable one. 
Two situations are considered. The first case is  in  the presence of the strong plasma density gradient. A lasting amplification is possible when a strong Langmuir wave, whose phase velocity matches with the x-ray group velocity, provides the density gradient along the way of the x-ray propagation. 
  The second case is  in the presence of a pre-excited Langmuir wave.   The author considers 
two lasers, with their frequency differing by the Langmuir wave,   counter-propagating and  exciting a rather strong Langmuir wave.    
In the presence of  the mentioned Langmuir wave,
it is demonstrated  that 
the  ponderomotive interaction  between the x-ray and one of the lasers
excites the plasma density perturbation with a desired phase, resulting to amplify (decay) the x-ray (visible light laser). The condition for achieving the mentioned favorable phase  and the amplification strength, when the condition is met, is estimated.   The amplification strength is quite high, where the gain per length could reach 1000 per cm in both cases. 

\end{abstract}


\maketitle


There have been growing interests for intense visible-light laser and x-ray light. A lot of practical applications have been proposed~\cite{cpa, cpa2, cpa3,cpa4, brs, brs2, brs3, Free, Free2,langdon, Tabak, Fisch, malkin1, Gallardo, monoelectron, ebeam}. 
However, unlike the visible light laser technologies, 
the x-ray technologies still need much more progresses for  many potential applications~\cite{soft, soft2, soft3,bio, bio2, litho, litho2, litho3, litho4}.  
If the comparison is made between  the x-ray light and the visible light lasers,
one key missing ingredient in the x-ray technology is a strong amplifier. 
The existence of a strong  amplifier   will stimulate the critical applications such as  the atomic spectroscopy~\cite{soft, soft2, soft3}, the dynamical imaging of fast biological processes~\cite{bio, bio2} and the next-generation semi-conductor lithography~\cite{litho, litho2, litho3, litho4}.
In this paper, 
one strong x-ray amplification by an intense visible-light laser is considered based upon the non-resonant Raman scattering between a   x-ray and a laser. 


The Raman scattering is very well-known non-linear laser-plasma interaction~\cite{Fisch, malkin1, sonbackward, BBRS, BBRS2, BBRS3}.   
 The ponderomotive interaction between two lasers excites \textbf{resonantly} a Langmuir wave.  This excited Langmuir wave rotates the energy  
from the higher frequency laser to the lower frequency laser.
The author is motivated by  the possibility that 
the Raman scattering might rotate the energy from 
a visible-light laser to a x-ray so that  the x-ray light gets amplified.  
In the conventional Raman scattering, however,   the energy flows from the higher frequency laser to the lower frequency laser, not the opposite way.  
 Furthermore, 
the ponderomotive interaction between a visible-light laser and a x-ray 
cannot excite a \textbf{resonant}  plasma wave   due to their frequency difference.  As well-known and further explained later in this paper, 
the phase of the  plasma  density perturbation from the resonant (non-resonant) scattering is  out of sync by a quarter cycle (in sync) with the ponderomotive interaction  and is suitable (not suitable) for the energy exchange.
These physical constraints seem to suggest that 
 the x-ray amplification from the non-resonant Raman scattering with the visible-light laser is  not feasible.

The author shows in this paper 
that, in certain cases, this undesirable phase can be manipulated for the x-ray amplification.  Two cases are considered. As the first case, in the presence of a strong density gradient, 
 the non-resonant ponderomotive density perturbation between a x-ray and a visible-light laser  could be induced with the suitable phase so that the x-ray can be amplified by a visible light laser. 
 It is also proposed that 
a co-propagating Langmuir wave can be  utilized to provide the lasting and strong density gradient by matching its phase velocity with the group velocity of the amplified x-ray.  
As the second case,
the author considers the situation 1)
 Two counter-propagating lasers excite a rather strong Langmuir wave and  
2) under a certain condition,  the ponderomotive interaction between  the lasers and the x-ray light generates the density perturbation with the desirable phase in the presence of the excited Langmuir wave so that  the x-ray amplification is possible.
The manipulation of the phase for the non-resonant ponderomotive density perturbation, which  is not possible under the normal circumstances,  is made possible by the presence of the excited strong Langmuir wave.  
The strength of the amplification is computed in this paper as a function of the laser intensity, the electron density, the strength of the excited Langmuir wave  and the x-ray frequency. 
The condition and the strength of the x-ray amplification is estimated as a mathematical function of the density gradient strength, the laser intensity and the x-ray frequency.  
The amplification strength could be very strong with the  gain-per-length  reaching 1000 per cm in both cases.  
   

\section{ Raman scattering and the energy transfer between the lasers}

Let us briefly review the backward Raman scattering physics. 
Consider two lasers with the frequency $\omega_1 $ and $\omega_2$ counter-propagating in the z-direction, where $\omega_1 > \omega_2$. For simplicity, let us assume that the lasers are linearly polarized and  their electric fields are parallel to each other. Their electric field is then given as 
$E_{x1} = E_1 \exp \left(\omega_1 t - k_1 z\right)$ and $E_{x2} = E_1 \exp \left(\omega_2 t + k_2 z\right)$.
Those lasers excite the density perturbation in the z-direction via the ponderomotive interaction.   
 The density response to the ponderomotive potential can be obtained from the continuity equation and the momentum equation: 

\begin{eqnarray}
 \frac{\partial \delta n_e }{\partial t}  &=&-  \mathbf{\nabla} \cdot ( \delta n_e \mathbf{\mathrm{v}} ) \nonumber \mathrm{,}\\ \nonumber \\
 m_e \frac{d \mathbf{v} }{dt} &=& e\left( \mathbf{\nabla} \phi  - \frac{\mathbf{v}}{c} \times \mathbf{B}\right) \nonumber \mathrm{,} \\ \nonumber
\end{eqnarray}
Combining the above equations with the Poisson equation $\nabla^2 \phi = -4 \pi \delta n_e e $, the density response is governed by~\cite{McKinstrie}

\begin{eqnarray} 
\left(\frac{\partial^2 }{ \partial t^2 } + \omega_{\mathrm{pe}}^2\right)\delta n_e &=&
\frac{en_0 }{m_ec} \mathbf{\nabla} \cdot \left( \mathbf{v} \times \mathbf{B} \right)\mathrm{.} \nonumber \\ \nonumber \\
&=&  -n_0 (ck_1 + ck_2)^2 a_1 a_2^*\mathrm{,}
 \label{eq:den}
\end{eqnarray} 
where $\omega_{\mathrm{pe}}^2= 4\pi n_0 e^2/ m_e$ is the plasma Langmuir wave frequency, $n_0$ is the background electron density,  all physical quantities are expressed as $b(z,t) = b \exp\left(i \omega t - k z\right) + b^*\exp\left(-i \omega t + k z\right) $, $a_{1,2} = eE_i/m\omega_{1,2} c $ is the laser quiver velocity normalized by the velocity of the light and 
$E_{i}$ is the electric field of the laser $i$.

The density perturbation from the ponderomotive force can be   estimated from Eq.~(\ref{eq:den}). With the frequency $\omega_{\mathrm{pe}} \neq  \omega_1 - \omega_2 $, it is given

\begin{eqnarray} 
\delta n(\omega_1 - \omega_2, k_1 + k_2)  &=& -n_0 \frac{(ck_1 + ck_2)^2 }{ (\omega_1-\omega_2)^2 - \omega_{\mathrm{pe}}^2} a_1^* a_2  \nonumber \\ \nonumber \\
&=& -n_0 C_B a_1^* a_2 \mathrm{,} \label{eq:nonresonance}
\end{eqnarray} 
where  $C_B = (ck_1 + ck_2)^2 /( (\omega_1-\omega_2)^2 - \omega_{\mathrm{pe}}^2) >1 $ is a positive real constant, because $\omega_{1,2} > \omega_{\mathrm{pe}}$.  
The lasers will respond to the density perturbation in Eq.(\ref{eq:den}) by~\cite{McKinstrie}: 

\begin{eqnarray}
L_1 a_1  &=& +i \frac{\omega_{\mathrm{pe}}^2 }{2\omega_1} \left( \frac{\delta n}{n_0} a_2\right)    \label{eq:1} \mathrm{,}\\ \nonumber \\ 
L_2 a_2  &=& +i \frac{\omega_{\mathrm{pe}}^2 }{2\omega_2} \left( \frac{\delta n^*}{n_0} a_1\right)    \label{eq:2} \mathrm{,}  
\end{eqnarray}
where $L_{i} = (\partial/\partial t) + v_{i} (\partial/\partial z) $ with $v_{i}$ is the group velocity of the laser. 
In case  $ \omega_1-\omega_2\neq \omega_{\mathrm{pe}}$, the density perturbation is in phase with the ponderomotive potential, as $\delta n \cong -C_Ba_1^* a_2 n_0$. Then,  
    $L_2 a_2  =  i(\omega_{\mathrm{pe}}^2 / 2 \omega_1) C_B|a_1|^2 a_2$; 
the energy $|a_2|^2$ does not change but only its phase  is modulated. However, in the case $ \omega_1-\omega_2 = \omega_{\mathrm{pe}}$, 
the resonance occurs and the density perturbation is given as~\cite{McKinstrie}

\begin{equation}
\delta n(t)  = +\mathbf{i}\left( \frac{(c k_1 + c k_2)^2 }{2\omega_{\mathrm{pe}}}   a_1 a_2^*  \right)t    \label{eq:resonance} \mathrm{,}
\end{equation}
Then, $\delta n\cong i C_B(t) a_1 a_2^*$, where $C_B(t)>0$ is the real number so that 
 $L_2 a_2  =   C_B(t)(\omega_{\mathrm{pe}}^2 / 2 \omega_2)|a_1|^2 a_2$ from  Eq.~(\ref{eq:2}). 
The energy of the laser $\omega_2$ grows while the energy of the laser $\omega_1$ decays. 
This density response, lagging by a quarter cycle (or imaginary response),  is the key for channeling energy from one laser to the other.  
The same process occurs in  a simple harmonic oscillator; 

\begin{equation}
\left(\frac{d^2  }{d t^2} +\omega_0^2\right) a  = A \cos(\omega t) \mathrm{.}
\end{equation}
If $\omega_0 \neq \omega$ ($\omega_0 = \omega$), then the particular solution is $a = A/(\omega_0^2 -\omega_1^2) \cos(\omega t) $ ($a = (A/\omega) t \sin(\omega t) $). The difference between  the cosine function and  the sine function is  what the imaginary number in Eq.~(\ref{eq:resonance}) represents. 
In the plasma where the x-ray light can propagate, noting that $\omega_1 > \omega_{\mathrm{pe}} $ and  
$ \omega_3-\omega_1 \gg \omega_{\mathrm{pe}} $, wherein  $\omega_3$ is the frequency of the x-ray light,  
  the energy transfer is not feasible  between the visible light and the x-ray light.  

\section{the electron density gradient and the phase-mismatch of the ponderomotive interaction}

In the previous section, it is shown that 
the non-resonant scattering cannot exchange the energy between the lasers.
 The plasma density gradient  can change the situation as will be shown.  
Consider a plasma with the density gradient $ n(x) = n_0 + (dn/dx) x $ and  an interaction of the x-ray $\omega_X$ with a counter-propagating visible light laser $\omega_1$. In this case, the Eq.~(\ref{eq:den}) will be given as  

\begin{eqnarray} 
\left(\frac{\partial^2 }{ \partial t^2 } + \omega_{\mathrm{pe}}(x)^2\right)\delta n_e =
\frac{e}{m_ec}\left((n(x) \mathbf{\nabla} \cdot  + \frac{dn}{dx} \right)\left( \mathbf{v} \times \mathbf{B} \right)\mathrm{.} \nonumber \\ \nonumber \\ \nonumber 
 = -\left(n(x) + i \frac{dn/dx}{k_X+k_1}\right)(ck_X + ck_1)^2 a_X a_1^*\mathrm{,} \\ 
+ \left(n(x) + i \frac{dn/dx}{k_X-k_1}\right)(ck_X - ck_1)^2 a_X a_1
\nonumber \\ 
 \label{eq:den1}
\end{eqnarray} 
Let us assume that $n(x)$ is slowly varying so that it is constant  to the x-ray.  
The first term and third terms on the right side of Eq.~(\ref{eq:den1}) is the normal ponderomotive potential and will have the real phase as $\omega_X - \omega_1 \gg \omega_{\mathrm{pe}}(x)$, while the second and fourth term has the imaginary phase-mismatch.  By using Eq.~(\ref{eq:den1}) in Eq.~(\ref{eq:2}), 
the following  can be obtained  

\begin{eqnarray} 
L_x a_X =  
  &-&\frac{ \omega_{\mathrm{pe}}(x)^2}{2\omega_X} \frac{dn/dx}{n(x)} \left( \frac{A}{ k_X + k_1}
+ \frac{B}{k_X - k_1}\right)|a_1|^2 a_X \mathrm{.}
\nonumber \\ \nonumber \\
&+&i\frac{ \omega_{\mathrm{pe}}(x)^2}{2\omega_X} 
\left(A + B \right)|a_1|^2 a_X   \label{eq:den3} \\ \nonumber
\end{eqnarray} 
where $A = (ck_X+ck_1)^2/ ( (\omega_X - \omega_1)^2 - \omega_{\mathrm{pe}}^2) \cong 1$ and  $ B = (ck_X-ck_1)^2/ ( (\omega_X + \omega_1)^2 - \omega_{\mathrm{pe}}^2) \cong 1 $ because $\omega_X \gg \omega_1$.  
The second term on the right side of Eq.~(\ref{eq:den3}) comes from the first and third terms on the right side of Eq.~(\ref{eq:den1}) and pure phase-modulation.  On the other hand, 
the first  term on the right side of Eq.~(\ref{eq:den3}) comes from the second term and fourth terms on the right side of Eq.~(\ref{eq:den1}), and this term due could rotate the energy from one E\&M field  to another becuase of its phase.

For the case of the co-propagating visible light laser, the same analysis and conclusion  can be made as given in Eq.~(\ref{eq:den3}) by replacing 
 $A = (ck_X-ck_1)^2/ ( (\omega_X - \omega_1)^2 - \omega_{\mathrm{pe}}^2) \cong 1$ and  $ B = (ck_X+ck_1)^2/ ( (\omega_X + \omega_1)^2 - \omega_{\mathrm{pe}}^2) \cong 1 $.   If $dx/dn < 0$, the amplification of x-ray occurs from the second term.  
In both cases, the growth rate can be estimated  $\Gamma \cong -(\omega_{\mathrm{pe}}^2/\omega_X)(dn/dx/k_X n)|a_1|^2 $. 


\section{The phase-mismatching through the density gradient of the plasma wave} 

While the plasma density gradient has a desired impact on the phase of the ponderomotive density perturbation, 
its amplification strength is limited by   the fact that $\alpha = (dn/dx/k_Xn_0) \ll 1$.   Furthermore, as the x-ray moves along, $dn/dx$  tends to change the sign fast and the average of $\alpha$ is even smaller. 
However, the author contemplates   a circumstance 
when  the density gradient is sustained  by a plasma Langmuir wave wave and thus  $\alpha$ being not so small. 
As demonstrated many times, the plasma can support a very strong plasma wave with the high electron density gradient. 
In addition, the particularly interesting  case  when   
the phase velocity of the Langmuir wave wave  is the same as the group velocity of the x-ray light.  In this case,   the x-ray will propagate experiencing the constant strong electron density gradient instead of the fast oscillating one. 
The x-ray sees the lasting and strong plasma density gradient so that it is strongly amplified.

A good candidate is the Langmuir wave excited via  the forward Raman scattering by two co-propagating lasers. 
The phase velocity of the wave is close to the velocity of the light. Even if the lasers do not satisfy the plasma resonance condition, a large amplitude density perturbation with the phase velocity close to the light will be good enough for our purpose.  
For now,  let us assume that a strong electron density perturbation exists inside the plasma, with the phase velocity the same with the x-ray group velocity.

\begin{figure}
\scalebox{0.3}{
\includegraphics{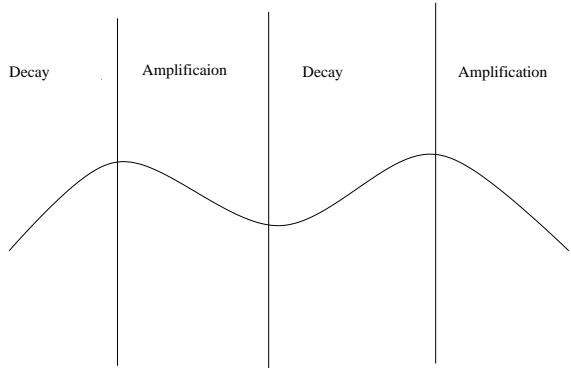}}
\caption{\label{fig:1}
The X-ray  amplifies (decays) in the region where $dn/dx<0$ ($dn/dx>0$).  
}
\end{figure}

 Let us denote the electron density as $n(z) = n_0 + n_L(z-v_Xt) $, where $n_L$ is non-negligible.  
The goal is to estimate the density response $\delta n $ to the ponderomotive interaction between a x-ray and an intense laser. 
Take the time derivative on the continuity equation, assuming the zeroth order of the density is  $n_0 + n_L(x-v_Xt)$, we obtain the linearized equation 

\begin{equation}
\frac{\partial^2 \delta n }{\partial t^2}=  - n(x) \nabla\cdot( \frac{\partial \mathbf{v}}{\partial t} )- \frac{dn_L}{dx}\frac{\partial v}{\partial t}  - \frac{\partial n_L}{\partial t}\frac{d v}{d x}  \mathrm{,} \label{eq:base}
\end{equation}
where the velocity $\mathbf{v} $ is generated by both the electro-static response and  the ponderomotive interaction between the x-ray $a_X(\omega_X, k_X)$ and an intense laser $a_1(\omega_1, \pm k_1)$. The first term on the right side of Eq.~(\ref{eq:base}) is the usual ponderomotive excitation and the Poisson contribution. Due to the functional form $n_L$, $\partial n_L / \partial t = -v_X dn_L/dx$.   If the velocity $\mathbf{v} $ is caused by a co-propagating laser, 
   $\mathbf{v} = \mathbf{v}( (\omega_X + \omega_1)t - (k_X + k_1)z)$ by the beating of $a_X a_1$ and  $d\mathbf{v}/dz = -c(\omega_X + \omega_1)/ (k_X + k_1) (\partial \mathbf{v}/ \partial t)$. 
Also,    $\mathbf{v} = \mathbf{v}((\omega_X - \omega_1)t - ( k_X - k_1)z)$ by the beating of $a_X a_1^*$ and   $d\mathbf{v}/dz = -c(\omega_X - \omega_1)/ (k_X - k_1) (\partial \mathbf{v}/ \partial t)$. Putting all these together,  we obtain 

\begin{eqnarray}
\frac{\partial^2 \delta n }{\partial t^2} &=& 
- n(x) \nabla\cdot( \frac{\partial \mathbf{v}}{\partial t} ) \nonumber \\
 \nonumber 
&-& \frac{dn_L}{dx}\left[1 + A   \right]\frac{\partial v(a_Xa_1)}{\partial t}   \\ 
&-& \frac{dn_L}{dx}\left[1 + B \right]\frac{\partial v(a_Xa_1^*)}{\partial t}   \mathrm{,} \label{eq:den4}
\end{eqnarray}
where $\omega_{\mathrm{pe}}^2 = 4 \pi n(x) e^2 /m_e$, 
$A =  v_X (k_X + k_1)/(\omega_X + \omega_1) $ and $B=  v_X (k_X - k_1)/(\omega_X - \omega_1) $. 
 For the counter propagating wave, the same can be derived by replacing 
$A =  v_X (k_X - k_1)/(\omega_X + \omega_1) $ and $B=  v_X (k_X + k_1)/(\omega_X - \omega_1) $.
 Because $\omega_X \gg \omega_1$ and $k_X \gg k_1$, $A\cong 1 $ and $B\cong 1$. 
By using Eqs.~(\ref{eq:1}) and  (\ref{eq:den4}), 
we obtain  the evolution of x-ray envelope as 

\begin{equation} 
L_x a_X =-\Gamma a_X=
  -\frac{ \omega_{\mathrm{pe}}(x)^2}{2\omega_X } \left(4 \frac{dn_L/dx}{k_X n(x)}  - 2 i\right)
   |a_1|^2 a_X \mathrm{.} \label{eq:xray}
\end{equation} 
Eq~(\ref{eq:xray}) is the major result of this paper. 
As the x-ray propagates experiencing the same phase of the plasma density perturbation, the x-ray could be amplified (decayed) via the laser energy in the region where $dn_L /dx< 0$ ($dn/dx>0$), as illustrated by Fig.~(\ref{fig:1})

As an example, consider $n_0 = 10^{20} \ \mathrm{cm}^3 $ and  $n_L = 0.3 \  n_0$ so that $\omega_{\mathrm{pe}} = 5.64 \times 10^{14} / \sec$ and  
$n_L \cong \exp(k_L (x- v_X)t) $ with $k_L \cong \omega_{\mathrm{pe}} / c = 1.9 
\times 10^{4} / \mathrm{cm}$. 
Considering the x-ray with the wave length of $0.2 \ \mu \mathrm{m}$, and a laser with    the wave length of $1 \ \mu \mathrm{m}$, we obtain from Eq.~(\ref{eq:xray}) 
\begin{equation}
\Gamma = 2 \times 10^{12} \ I_{16} / \sec \mathrm{,}  \label{eq:ex}
\end{equation}  
where $I_{16} $ is the laser intensity normalized by $ 10^{16} \  W/\mathrm{cm}^2$. From  Eq.~(\ref{eq:ex}),  it can be estimated that  the gain-per-length could reach 1000 per centimeter if $I_{16} \cong 10 $. The necessary minimum plasma size is $ L > c / \Gamma\cong (0.015/I_{16}) \ \mathrm{cm}$. 
For the same plasma and laser, but with the x-ray with  the wave length of $0.1 \ \mu \mathrm{m}$,   $\Gamma =5 \times  10^{11} I_{16} \ / \sec$ and $L > c / \Gamma\cong (0.05/I_{16}) \  \mathrm{cm}$. 
By having the interaction length large enough, the x-ray could be amplified by many factors. 

If the electron density is low or the laser intensity is weak, 
the size of the plasma for the amplification needs to be larger, inversely proportional to the electron density, laser intensity and the strength of the Langmuir wave. 
As for the excitation of the density perturbation or the Langmuir wave, 
 the key is to make the phase velocity of the wave to be the same with the group velocity of the x-ray. 
In the case when the forward Raman scattering is utilized, 
the laser has $\omega = \sqrt{\omega_{\mathrm{pe}}^2 + c^2k^2} > ck $ so the phase velocity $\delta \omega / \delta k > c$. However, it is possible to make 
$(\omega_1 - \omega_2)/|\mathbf{k}_1-\mathbf{k}_2| = v_X < c$ by injecting two lasers in a slightly skewed direction. 
 It is worthwhile to note that 
 in normal circumstances, the mechanism described is a weak amplification as the gradient oscillates the sign along the x-ray propagation. By matching the phase velocity of the wave  with the group velocity of the x-ray, 
this oscillation is suppressed and its amplification could be strong.

\section{X-ray amplification in the presence of the pre-existing Langmuir wave}

\begin{figure}
\scalebox{0.4}{
\includegraphics{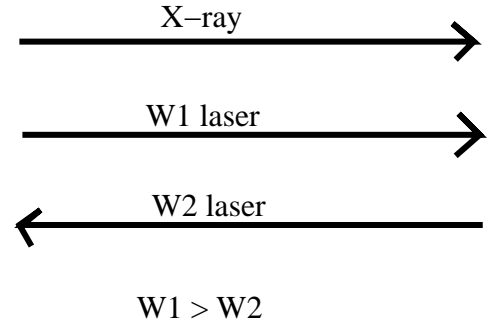}}
\caption{\label{fig:2}
The direction of the X-ray and the two lasers.
}
\end{figure}

Consider two lasers with the frequency $\omega_1 $ and $\omega_2$ counter-propagating in the z-direction, where $\omega_1- \omega_2 =\omega_{\mathrm{pe}}$.  The density perturbation $\delta n$ will respond to the beating ponderomotive potential as given in Eq.~(\ref{eq:resonance}). 
As the density perturbation is considerable,  the assumption that $\delta n \ll n_0 $ might not be valid any more, 
in which situation  the zeroth order is  $n = n_0 + \delta n  = n_0 + n_L(\omega_{\mathrm{pe}}, k_1+k_2) \cong n_0 + \mathbf{i} C_B a_1 a_2^*$. As shown later, the imaginary response of the $n_L \cong \mathbf{i} C_B a_1 a_2^* $ is the key ingredient of the analysis in this paper. 
In the presence of $n_L$, the author computes  
the pondermotive density response between the x-ray and the lasers. 
Denote a x-ray light frequency  as  $\omega_3$ and the wave vector as $k_3$, travelling in the same direction with $\omega_1$ laser. See Fig.~(\ref{fig:2}) for the directions of lasers and the x-ray light. 
The expansion of the continuity equation  is given as


\begin{eqnarray}
\left(\frac{\partial^2  }{\partial t^2} + \omega_{\mathrm{pe}}^2\right)\delta n_e  &=&-  \mathbf{\nabla} \cdot ( \frac{\partial n_L}{\partial t} \mathbf{v} ) -  \mathbf{\nabla} \cdot ( (n_0+n_L)\frac{\partial \mathbf{v}}{\partial t} ) )
\nonumber \\ \nonumber \\
 &\cong&   -  (n_0 + n_L) \mathbf{\nabla} \cdot (\frac{\partial \mathbf{v}}{\partial t}) -  \left(\mathbf{\nabla} n_L\right)\cdot  (\frac{\partial \mathbf{v}}{\partial t} ) \mathrm{.} \nonumber \\ \nonumber \\ \label{eq:exp}  
\end{eqnarray}
The term involving $ (\partial n_L/\partial t) $ in Eq.(\ref{eq:exp}) is smaller than other terms by $\omega_{\mathrm{pe}}/\omega_3 \gg 1$, which will be ignored.  Then, there will be three remaining terms on the right side of Eq.~(\ref{eq:exp}).  The first term involving $n_0$ is  the usual  ponderomotive density response  between the x-ray and the lasers without the preexisting Langmuir wave. 
The second term with $n_L$ and the third term with $\nabla (n_L)$ is the density response due to the presence of the Langmuir wave.


In the above equation,
the velocity $\mathbf{v}$ is excited by the ponderomotive force  between one of the laser and the x-ray.
There are four ponderomotive interactions ($a_3 a_1 $, $a_3 a_1^*$,  $a_3 a_2 $, $a_3 a_2^*$), which could interact with $n_L$ or $n_L^*$. Then, 
there are eight density perturbations, ($n_La_3 a_1 $, $n_La_3 a_1^*$,  $n_La_3 a_2 $, $n_La_3 a_2^*$, $n_L^*a_3 a_1 $, $n_L^*a_3 a_1^*$,  $n_L^*a_3 a_2 $, $n_L^*a_3 a_2^*$). Those eight density perturbations beat with the laser quiver ($a_1$, $a_1^*$, $a_2$, $a_2^*$). 
Then, there are 32 combinations  and the relevant ones for our interest  
 is the ones with the total total momentum $k_3$ and the frequency $\omega_3$;
There are four such combinations $(n_L [a_1 a_3] a_2^*, n_L [a_2^* a_3] a_1, 
n_L^* [a_1^* a_3] a_2, n_L [a_2 a_3] a_1^*)$, where
$n_L [a_1 a_3] a_2^*$, for an example, represents  a density perturbation, induced by a Langmuir wave $n_L$ and  the ponderomotive force by $a_1 a_3$,  make a x-ray beat current with the laser quiver $a_2^*$.

With the above consideration in mind, by computing all four terms, 
the relevant density response for the second term in Eq.~(\ref{eq:exp}) is given as 
\begin{eqnarray}
\delta n_1 = &+& in_0\left(+ A_1 |a_1|^2 a_2 - A_2 |a_2|^2 a_1^* \right) a_3 \nonumber \\   \nonumber \\ 
 &+&in_0\left( A_3 |a_1|^2 a_2^* - A_4 |a_2|^2 a_1 \right)a_3 \mathrm{,} \label{eq:first}
\end{eqnarray}
where $A_1 = C (ck_3 + ck_1)^2/(\omega_3+\omega_2)^2$ represents 
the ponderomotive potential by $a_3 a_1 $, producing the x-ray current 
by beating with the $a_2^*$ quiver and  $n_L^*$,
 $A_2 =C (ck_3 + ck_2)^2/(\omega_3-\omega_1)^2$ represents
the ponderomotive interaction  by  $a_3 a_2^* $ producing the x-ray current 
by  beating with the $a_1$ quiver and  $n_L^*$. 
$A_3 =C (ck_3 - ck_1)^2/(\omega_3-\omega_2)^2$ represents 
the ponderomotive potential by $a_3 a_1^* $ producing x-ray current by beating with the $a_2$ quiver  and $n_L$. 
 $A_4 =C (ck_3 - ck_2)^2/(\omega_1+\omega_3)^2$ represents 
the ponderomotive potential being excited by  $a_3 a_2 $ producing 
the x-ray current by beating with the $a_1^*$ quiver and  $n_L$.

The  relevant density response for the third term in Eq.(\ref{eq:exp}) can be computed in a similar fashion and the final form similar to Eq.(\ref{eq:first}) is given as 
\begin{eqnarray}
\delta n_2 &=& + in_0\left(B_1 |a_1|^2 a_2 - B_2 |a_2|^2 a_1^* \right)  \nonumber \\   \nonumber \\ 
 &+& in_0\left( -B_3 |a_1|^2 a_2^* + B_4 |a_2|^2 a_1 \right)a_3 \mathrm{,} \label{eq:second}
\end{eqnarray}
where $B_1 = C (ck_3 + ck_1)(ck_1+ck_2)/(\omega_3+\omega_2)^2$ represents
the ponderomotive potential by $a_3 a_1 $ producing the x-ray current 
by beating with the $a_2^*$ quiver and beating with $n_L^*$.
 $B_2 =C (ck_3 + ck_2)(ck_1+ck_2)/(\omega_3-\omega_1)^2$ represents the ponderomotive potential by  $a_3 a_2^* $ producing the x-ray current by  beating with the $a_1$ quiver and  $n_L^*$,
$B_3 =C (ck_3 - ck_1)(ck_1+ck_2)/(\omega_3-\omega_2)^2$ represents 
the ponderomotive potential by $a_3 a_1^* $ producing the x-ray current by beating with the $a_2$ quiver and  $n_L$ and 
 $B_4 =C (ck_3 - ck_2)(ck_1+ck_2)/(\omega_1+\omega_3)^2$ represents the ponderomotive potential by  $a_3 a_2 $ producing the x-ray current by beating with the $a_1^*$ quiver and $n_L$.

Combining  Eq.~(\ref{eq:first}) and Eq.~(\ref{eq:second}) with Eq.~(\ref{eq:1}) or Eq.~(\ref{eq:2}), 
we obtains  

\begin{eqnarray}
 L_3 a_3  &=&  \frac{\omega_{\mathrm{pe}}^2 }{2\omega_3} \left( 
\Gamma C_B |a_1|^2 |a_2|^2\right)a_3 \nonumber \\ \nonumber \\
&=&    \frac{\omega_{\mathrm{pe}}^2 }{2\omega_3} \left(\frac{n_L}{n_0}\right)\left( \Gamma  |a_1| |a_2|\right)a_3 \label{eq:major} \\ \nonumber
\end{eqnarray}
where $ \Gamma = A_1 +A_2 + B_1 + B_2 - A_3-A_4 - B_3 - B_4$
and $|n_L| \cong C_B |a_1a_2^*|$. 
The equation~\ref{eq:major} is the major result of this paper. 
It can be also shown that $\Gamma < 0 $ if $\omega_1 < \omega_2$.

As an example, consider $\omega_1 = 4 \ \omega_{\mathrm{pe}} $, $\omega_2= 3 \ \omega_{\mathrm{pe}}$ and   $\omega_3= 12 \ \omega_{\mathrm{pe}}$.
Then, 
$A_1 = 1.14,  A_2 =3.52,  A_3 =0.79, A_4=0.32$,
$B_1 = 0.50,  B_2 =1.64,  B_3 =0.69, B_4=0.25$
and $\Gamma =  4.75$. 
As another example,  consider $\omega_1 = 4 \ \omega_{\mathrm{pe}} $, $\omega_2= 3 \ \omega_{\mathrm{pe}}$ and   $\omega_3= 30 \ \omega_{\mathrm{pe}}$.
Then $\Gamma =  0.91$.
As another example,  consider $\omega_1 = 4 \ \omega_{\mathrm{pe}} $, $\omega_2= 3 \ \omega_{\mathrm{pe}}$ and   $\omega_3= 100 \ \omega_{\mathrm{pe}}$.
Then, $\Gamma =  0.24$.
The more frequency difference between the x-ray light and the lasers, the coupling  for   amplification becomes weaker. 
If $\omega_1 < \omega_2 $, then  $\Gamma $ will be negative and the x-ray will be  decayed. 

For the example with the specific parameter, consider a plasma with $n_0 = 10^{20}  \ / \mathrm{cm}^3$,  $n_L / n_0 =0.1$, two lasers with $1 \ \mu \mathrm{m} $ and  $1.33 \ \mu \mathrm{m} $ and x-ray light of $0.33\ \mu \mathrm{m}$. 
Then, it can be estimated as $\Gamma  \cong 1.5 \times 10^{13} \sqrt{I^1_{18}I^2_{18} } \sec$, where $I^1_{18}$ ($I^2_{18}$) is the intensity of $a_1$ ($a_2$) laser normalized by $10^{18} \ W /\mathrm{cm}^2$. For the relativistic intensity, the gain-per-length could reach 1000 per centimeter. 
As an another example, consider the same  plasma and the same lasers but x-ray light  of $0.1\ \mu \mathrm{m}$ and  $n_L / n_0 =0.3$.
Then, it can be estimated as $\Gamma  \cong 3 \times 10^{12} \sqrt{I^1_{18}I^2_{18} } \sec$.
 For the relativistic intensity, the gain-per-length could reach 100 per centimeter.
As an another example, consider the same  plasma and the same lasers but x-ray light  of $0.03\ \mu \mathrm{m}$ and  $n_L / n_0 =0.3$.
Then, it can be estimated as $\Gamma  \cong 2 \times 10^{11} \sqrt{I^1_{18}I^2_{18} } \sec$.
 For the relativistic intensity, the gain-per-length could reach 10 per centimeter.


\section{Summary}

In this paper, the author shows that there could be two cases in which the intense visible light laser can amplify the x-ray via the Raman scattering.  
As the first case, 
it is shown  that the strong density gradient  could create a plasma condition in which an intense  visible light laser can amplify a x-ray light. 
 The ponderomotive interaction between  the laser and the x-ray 
generates the plasma density perturbation with the desired phase mismatch and 
  induces the x-ray amplification.  
The strong and lasting density gradient can be  provided by a Langmuir wave, co-propagating with the phase velocity close to the x-ray group velocity.
If optimized as prescribed in this paper,   the gain could be as high as 100 per centimeter (1000 per centimeter) for the soft x-ray (the ultra-violet light).

As the second case,   if there is an existing Langmuir wave excited by two lasers, 
 the density response due to the beating of the ponderomotive interaction (between the x-ray light and the lasers) and the Langmuir wave
could have a phase beneficial to the   x-ray amplification.  
The gain per length can be as high as 1000 per centimeter in this case.
there is a problem of premature energy transfer between the visible light lasers  through the BRS;   the energy is transferred from the higher frequency laser to the lower one, depleting the higher frequency laser, along the x-ray propagation.  
There could be a few ways to overcome this. One
 possible way is to inject another laser in the same direction with the higher-frequency laser in order to prevent the decay.
Another way, probably more efficient way, is to inject two lasers in a slightly skewed direction so that the portion of the higher frequency laser that amplifies the x-ray will be replaced gradually before it decays due to the BRS.


\bibliography{tera2}

\end{document}